\documentclass{article}
\usepackage{latexsym}
\usepackage{color}
\usepackage{amsmath}
\usepackage{epsfig}
\usepackage{caption}
\usepackage{graphicx}

\linethickness{0.6pt}

\title{Spontaneous Topological Locking and Symmetry Restoration of Meron Lattices in Synthetic Antiferromagnets}

\author{G\"ul\c{s}en Do\u{g}an$^1$, \"Umit Ak\i nc\i $^{2,*}$}
\date{} 

\begin{document}
\maketitle

\begin{center}
$^1$\textit{The Graduate School of Natural and Applied Sciences, Dokuz Eyl\"ul University, TR-35160 \.Izmir, Turkey}\\
$^2$\textit{Department of Physics, Dokuz Eyl\"ul University, TR-35160 \.Izmir, Turkey}\\
$^*$\textit{Corresponding author: umit.akinci@deu.edu.tr}
\end{center}

\vspace{0.5cm}


\vspace{0.5cm}

\section{Abstract}

Synthetic antiferromagnets offer a robust platform for stabilizing fractional topological textures, effectively circumventing the limitations of ferromagnetic systems. In this study, we utilize large-scale Monte Carlo simulations to investigate the spontaneous topological locking and structural symmetry restoration of meron-antimeron crystals within SAF bilayers subjected to easy-plane magnetic anisotropy. In the uncoupled monolayer limit, increasing anisotropy induces an extreme core-shrinking effect that physically expands the inter-core distance and triggers a $C_4 \rightarrow C_2$ symmetry breaking. However, the introduction of an ultra-weak interlayer antiferromagnetic exchange acts as an active structural scaffold. For rigid crystals, this coupling strictly enforces vertical synchronization, forming robust antiferromagnetic bimeron dipoles and fully restoring the macroscopic $C_4$ rotational symmetry. Furthermore, in highly expanded, pre-collapse crystals, we observe an anomalous interlayer-induced lattice compression that actively maximizes the exchange energy. At extreme anisotropy limits where macroscopic crystalline order irrecoverably collapses, the bilayer coupling continues to enforce a strict local topological locking of surviving isolated defects. These findings reveal a fundamental decoupling between local vertical synchronization and global structural order, providing a comprehensive theoretical roadmap for stabilizing and manipulating fractional topological textures in beyond-skyrmion spintronic architectures.

\label{sec:intro}
\section{Introduction}

The field of topological spintronics has witnessed a paradigm shift with the discovery of magnetic skyrmions and related spin textures, characterized by their non-trivial topology and particle-like stability \cite{ref_1, ref_2}. These structures are regarded as ideal information carriers for next-generation racetrack memory and logic-in-memory architectures due to their small size and the low current densities required for their manipulation \cite{ref_3, ref_20}. However, a significant bottleneck in the implementation of ferromagnetic skyrmions is the Skyrmion Hall Effect (SkHE), where the Magnus force causes a transverse deflection of the texture during current-driven motion, potentially leading to information loss at track boundaries \cite{ref_4}. Furthermore, the inherent stray fields in FM systems limit the packing density of topological bits.

To circumvent these limitations, research has increasingly turned towards antiferromagnetic systems, which offer zero net magnetization, immunity to external magnetic perturbations, and the complete suppression of the SkHE \cite{ref_5, ref_6, ref_barker}. Synthetic antiferromagnets (SAFs) have emerged as a versatile alternative, where two FM layers are coupled antiferromagnetically through a non-magnetic spacer via the Ruderman-Kittel-Kasuya-Yosida (RKKY) interaction \cite{ref_parkin, ref_7}. SAFs effectively bridge the gap by combining the experimental accessibility of FM materials with the desirable physics of antiferromagnetic systems, allowing for the stabilization of ``topological dipoles'' that cancel out the Magnus force \cite{ref_8, ref_9,ref_10}. Furthermore, recent studies demonstrate that the precise tuning of interlayer exchange in SAF systems can strongly influence the morphology and vertical locking of these topological textures \cite{ref_juge}.

While the majority of SAF-based research has focused on systems with perpendicular magnetic anisotropy  hosting skyrmions ($Q=1$), recent experimental breakthroughs in two-dimensional van der Waals magnets \cite{ref_tan} and structurally engineered thin films governed by shape/thickness anisotropy \cite{ref_jani} have redirected attention toward systems with easy-plane anisotropy ($K<0$). These in-plane magnetized environments naturally favor the stabilization of merons, antimerons, and their bound pairs (bimerons), which carry fractional topological charges ($Q=\pm 1/2$) and represent highly promising ``beyond-skyrmion'' candidates for next-generation devices \cite{ref_11, ref_gobel_review}. In particular, antiferromagnetic bimerons and in-plane topological dipoles exhibit advantageous stability and driving dynamics, allowing for faster and more efficient current-driven motion compared to traditional skyrmions \cite{ref_zhang_bimeron, ref_shen}. In monolayer chiral magnets, meron-antimeron crystals (MAXs) are particularly susceptible to this easy-plane anisotropy; as the anisotropy strength increases, the ideal $C_4$ symmetry of the macroscopic crystal often undergoes significant distortion, leading to an anisotropic $C_2$ symmetry or a complete structural decay into fragmented phases \cite{ref_gao, ref_gobel,ref_q}. The stability and spatial ordering of these fractional textures in SAF architectures, specifically under the influence of strong easy-plane anisotropy, remain largely unexplored. A fundamental question arises: can the interlayer exchange coupling not merely align these fractional textures, but actively ``rescue'' and restore the lost global symmetry of the monolayer crystal?

In this work, we employ large-scale Monte Carlo simulations to investigate the spontaneous topological locking and symmetry restoration of MAXs in SAF bilayers. Our results reveal that the interlayer antiferromagnetic exchange ($J$) triggers a sharp, first-order-like ``topological locking'' transition at an exceptionally weak coupling regime. Crucially, we demonstrate that this locking mechanism acts as a precise structural scaffold: for rigid crystals, it strictly restores the ideal $C_4$ macroscopic symmetry that is otherwise distorted in the monolayer limit. Furthermore, for highly expanded and ``softened'' crystals nearing structural instability, we reveal an anomalous interlayer-induced lattice compression. Finally, we define the physical boundaries of this ``topological rescue'' mechanism at extreme anisotropy strengths. We show that while the macroscopic crystalline order irrecoverably collapses, a sufficiently strong SAF coupling continues to enforce a strict local topological locking of the surviving defects, revealing a fundamental decoupling between local vertical synchronization and global structural order.

The remainder of this paper is organized as follows. In Sec.~\ref{sec:model}, we introduce the theoretical Hamiltonian and detail the Monte Carlo methodology. In Sec.~\ref{sec:results}, we present our numerical findings, focusing on the dual nature of spontaneous locking and structural symmetry restoration. Finally, in Sec.~\ref{sec:conclusion}, we summarize our conclusions and discuss the implications of our results for future spintronic applications.

\section{Model and Methodology}
\label{sec:model}

\subsection{Hamiltonian and Physical Model}

To investigate the topological locking mechanism and symmetry restoration in bilayer SAF systems, we employ a classical Heisenberg model on an $L \times L \times 2$ square lattice. The total energy of the bilayer system is governed by the Hamiltonian:

\begin{equation} \label{eq_1}
\begin{aligned}
\mathcal{H} = &-J_{intra} \sum_{\alpha=1}^{2} \sum_{\langle i,j \rangle} \vec{S}_{i}^{(\alpha)} \cdot \vec{S}_{j}^{(\alpha)} - J_{inter} \sum_{i} \vec{S}_{i}^{(1)} \cdot \vec{S}_{i}^{(2)} \\
&- \sum_{\alpha=1}^{2} \sum_{\langle i,j \rangle} \vec{D}_{ij}^{(\alpha)} \cdot (\vec{S}_{i}^{(\alpha)} \times \vec{S}_{j}^{(\alpha)}) - K_{ani} \sum_{\alpha=1}^{2} \sum_{i} \left(S_{i}^{(\alpha)z}\right)^2
\end{aligned}
\end{equation}
where $\vec{S}_{i}^{(\alpha)}$ is a unit spin vector ($|\vec{S}_{i}^{(\alpha)}| = 1$) at site $i$ of layer $\alpha \in \{1,2\}$. The first term represents the intralayer ferromagnetic exchange ($J_{intra} > 0$), which serves as the fundamental energy scale of the system. The second term describes the interlayer antiferromagnetic coupling ($J_{inter} < 0$) between the two layers, which is the primary driver for the formation of topological dipoles in SAF architectures \cite{ref_7, ref_8}. 
The third term accounts for the Dzyaloshinskii-Moriya interaction (DMI), where the DMI vector is defined as $\vec{D}_{ij}^{(\alpha)} = D \hat{r}_{ij}$. This choice of DMI orientation favors the nucleation of Bloch-type spin textures \cite{ref_12, ref_19}. The last term $K_{ani}$ represents the easy-plane magnetic anisotropy ($K_{ani} < 0$), which penalizes out-of-plane spin components ($S^z$) and facilitates the stabilization of MAXs. Note that $\langle i,j \rangle$ denotes summation over the nearest neighbors on the lattice.

\subsection{Topological Characterization and Observables}

To distinguish the emergent magnetic phases, we calculate the static spin structure factor for the in-plane components:

\begin{equation}
\label{eq_2}
S_{\perp}^{\alpha}(\vec{q}) = \frac{1}{N} \left\langle \left| \sum_{\vec{r}} S_{\vec{r}}^{(\alpha)x} e^{-i\vec{q}\cdot\vec{r}} \right|^2 + \left| \sum_{\vec{r}} S_{\vec{r}}^{(\alpha)y} e^{-i\vec{q}\cdot\vec{r}} \right|^2 \right\rangle
\end{equation}
where $N = L^2$ is the number of sites per layer. For the topological analysis, the local topological charge density $q_{i}^\alpha$ is computed using the Berg-Lüscher method \cite{ref_14}, which provides a geometrically consistent definition for discrete spin systems. By triangulating each lattice plaquette into two triangles ($i,j,k$), the charge is defined via the solid angle subtended by the three spins:

\begin{equation}\label{eq_3}
q_i^\alpha = \frac{1}{2\pi} \sum_{\epsilon=\pm 1} \arctan \frac{\vec{S}_i^\alpha \cdot (\vec{S}_j^\alpha \times \vec{S}_k^\alpha)}{1 + \vec{S}_i^\alpha \cdot \vec{S}_j^\alpha + \vec{S}_j^\alpha \cdot \vec{S}_k^\alpha + \vec{S}_k^\alpha \cdot \vec{S}_i^\alpha},
\end{equation}
where the indices $j \equiv i + \epsilon\hat{x}$ and $k \equiv i + \epsilon\hat{y}$ denote the nearest-neighbor sites, with $\hat{x}$ and $\hat{y}$ being the unit vectors of the square lattice.

The total topological charge $Q$ is obtained by the summation over the lattice. A localized sum of $|Q| \approx 1/2$ in the topological defect region  confirms the presence of a meron or antimeron. The degree of vertical synchronization and topological alignment between the two layers is quantified through the interlayer topological charge correlation function:

\begin{equation}
\label{eq_4}
\chi = \frac{\left\langle \sum_{i}q_{i}^{(1)}q_{i}^{(2)} \right\rangle}{\sqrt{\left\langle \sum_{i}(q_{i}^{(1)})^{2} \right\rangle \left\langle \sum_{i}(q_{i}^{(2)})^{2} \right\rangle}}
\end{equation}
where $\chi \to -1$ indicates a perfect spontaneous antiferromagnetic locking of the topological cores.

\subsection{Monte Carlo Simulation Details}

The numerical exploration of the phase space was performed using the standard Metropolis Monte Carlo (MC) algorithm \cite{ref_13}. To ensure the identification of the global energy minimum and to minimize trapping in metastable states, we implemented a simulated annealing protocol. The system was cooled from a high-temperature paramagnetic state ($T = 2.01$) down to the target temperature with a fine temperature step of $dT = 0.01$. The final temperature is $T=0.01$.

All physical parameters are scaled by the intralayer ferromagnetic exchange interaction as $J = J_{inter}/J_{intra}, d = D/J_{intra}, K = K_{ani}/J_{intra}, T = k_B T_{real}/J_{intra}$. The intralayer interaction is choosen  as ferromagnetic $J_{intra}=1.0$. To maintain the system in the regime of stable chiral textures, the dimensionless DMI strength was fixed at $d = 2.0$. For each parameter set $(K, J)$, the system was first thermalized for $5\times 10^5$ MC steps (MCS) per spin, followed by $5\times 10^5$ MCS for statistical averaging. We examined various lattice sizes up to $L=120$ to account for finite-size effects and to confirm the spontaneous nature of the locking transition.

\section{Results and Discussion}
\label{sec:results}

In this section, we present the results of our Monte Carlo simulations, focusing on the interplay between easy-plane anisotropy ($K$) and interlayer exchange coupling ($J$). To establish a robust baseline, we first analyze the monolayer limit ($J = 0$), where the structural integrity of the MAX is solely dictated by the competition between DMI and anisotropy. Subsequently, we investigate the bilayer SAF architecture to demonstrate the spontaneous topological locking mechanism and its role in restoring topological phase crystal symmetry. Unless otherwise specified, the results discussed herein correspond to a representative lattice size of $L=80$, which we found to be the optimal window for capturing topological textures without significant finite-size artifacts, as confirmed by our scaling analysis up to $L=120$.

\subsection{Phase Evolution in the Monolayer Limit ($J=0$)}
\label{subsec:monolayer}

\begin{figure}[!htbp]
    \centering
\includegraphics[width=7.4cm,height=7cm]{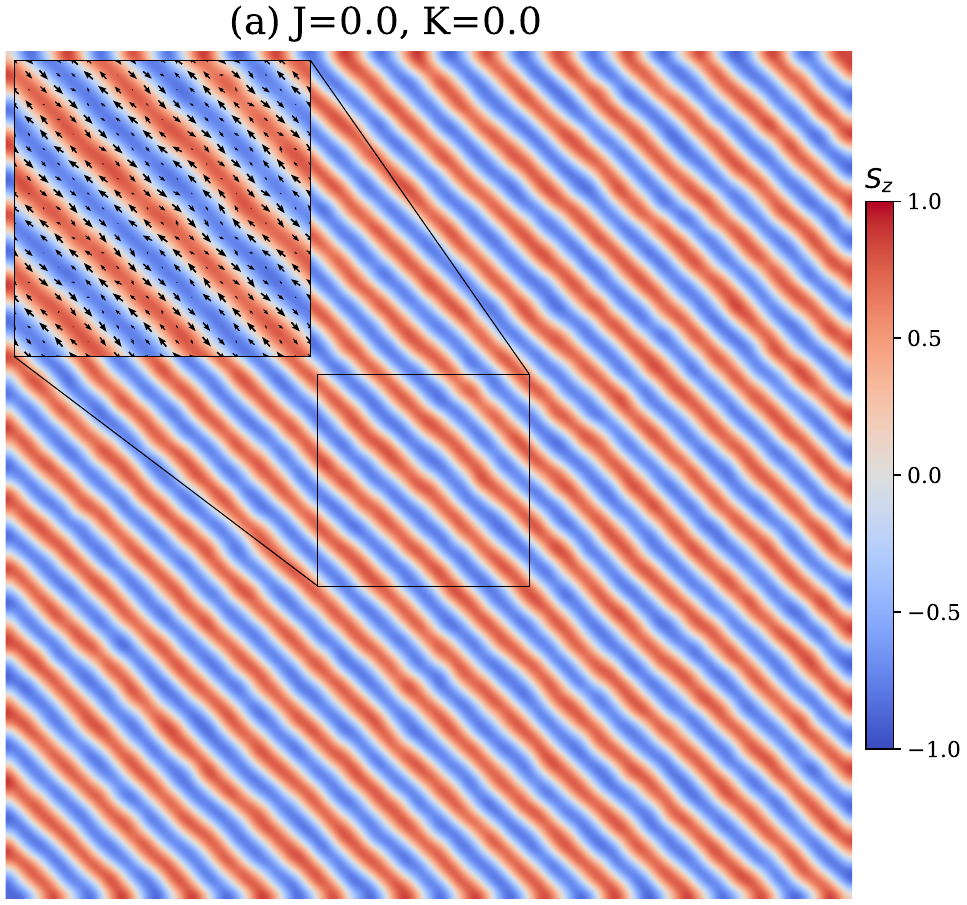}
\includegraphics[width=7.4cm,height=7cm]{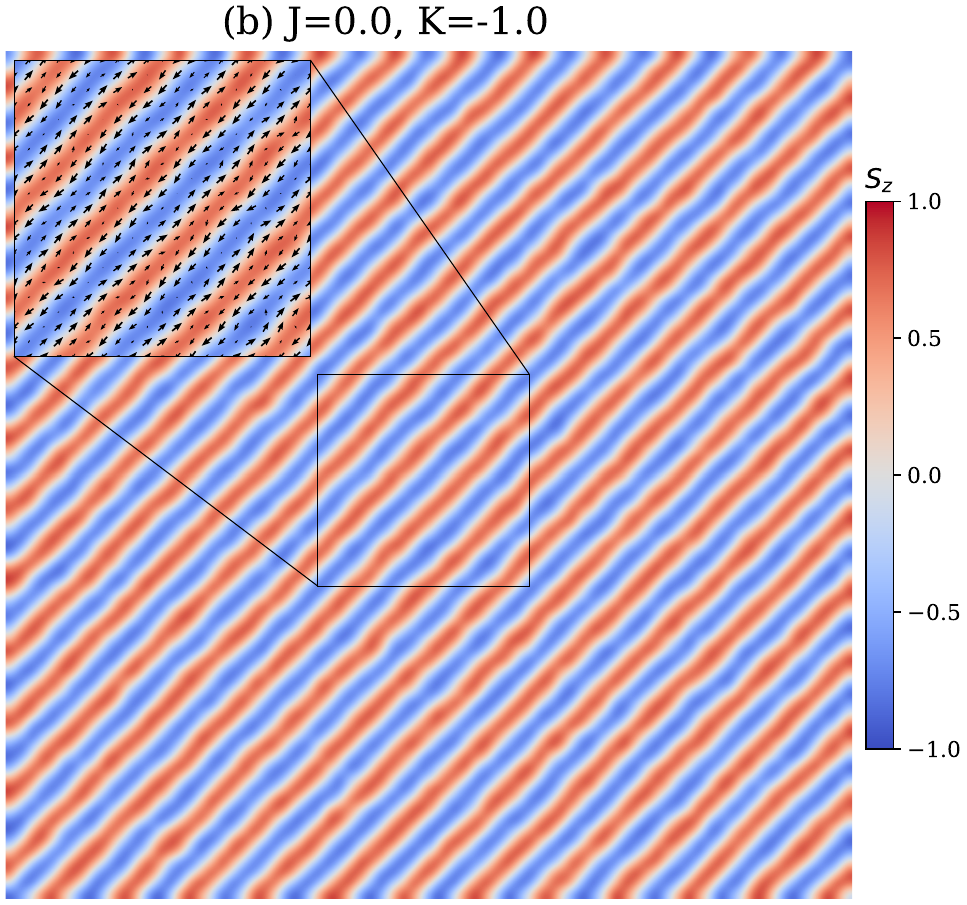}

\includegraphics[width=7.4cm,height=7cm]{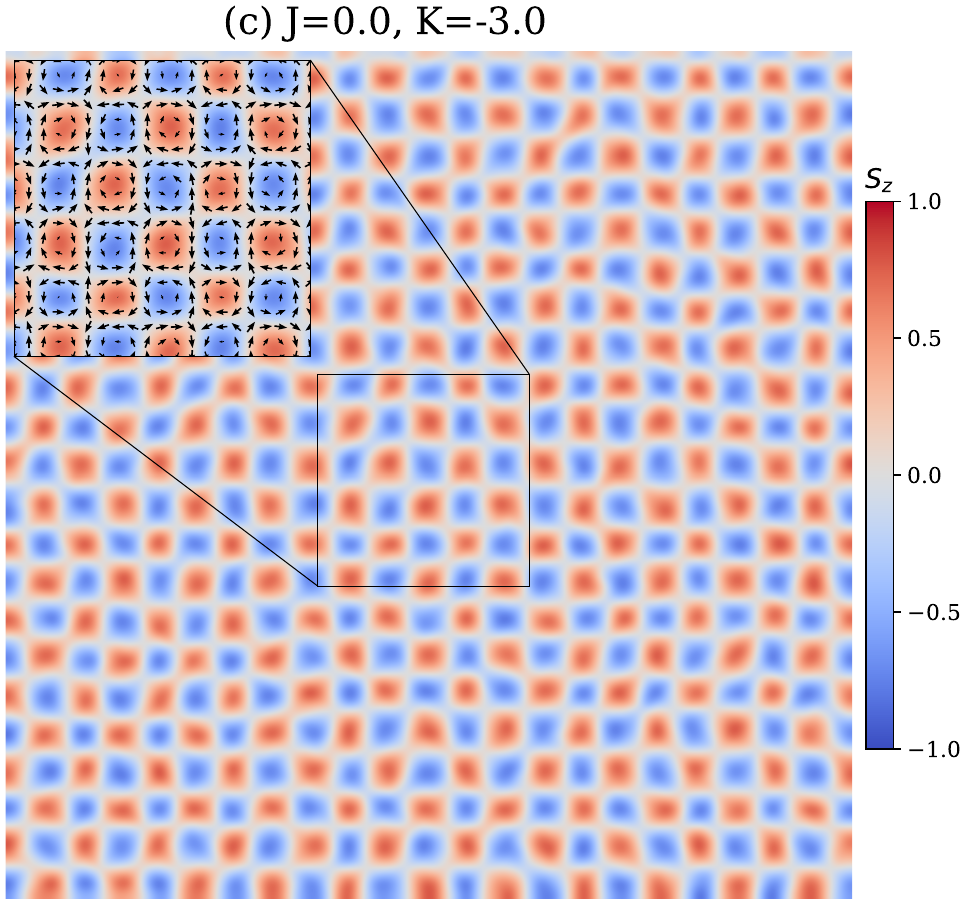}
\includegraphics[width=7.4cm,height=7cm]{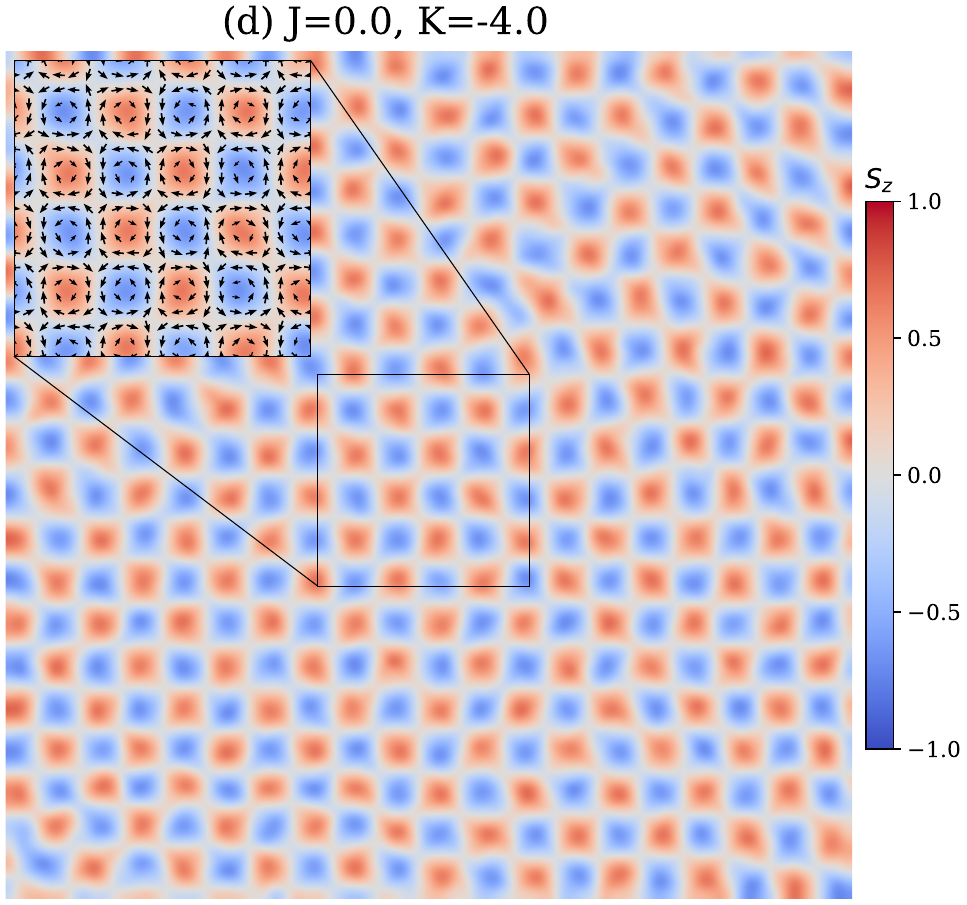}

\includegraphics[width=7.4cm,height=7cm]{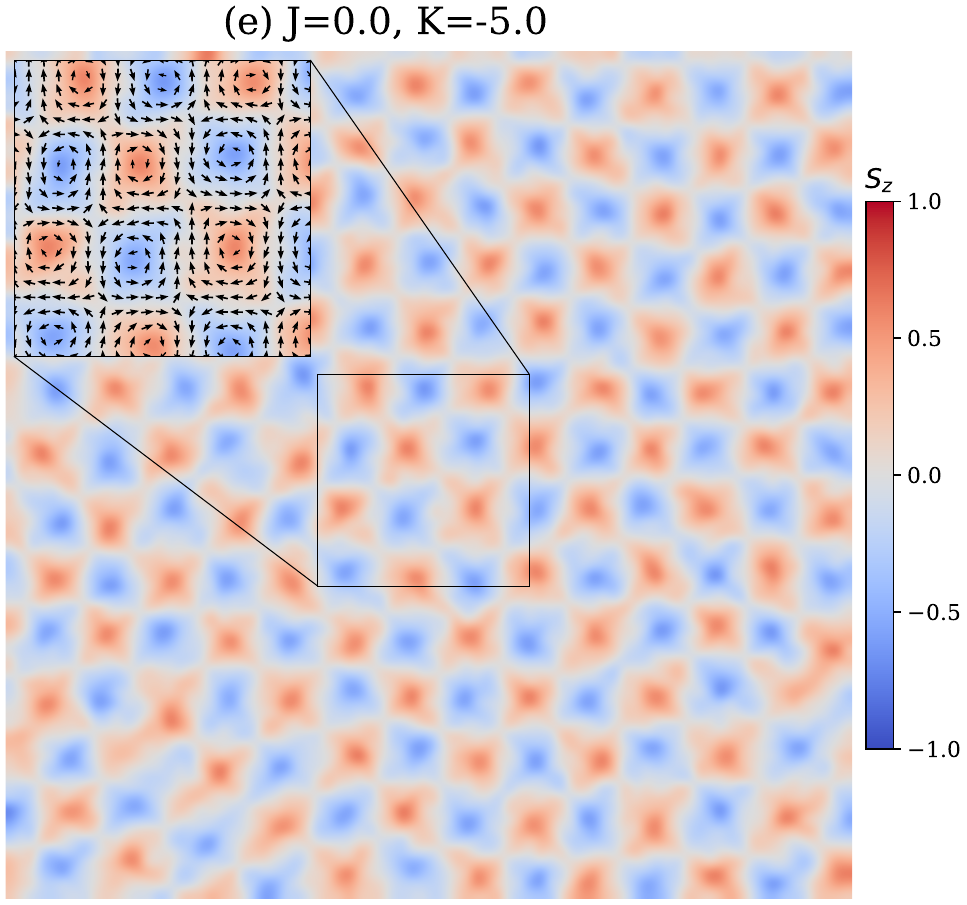}\includegraphics[width=7.4cm,height=7cm]{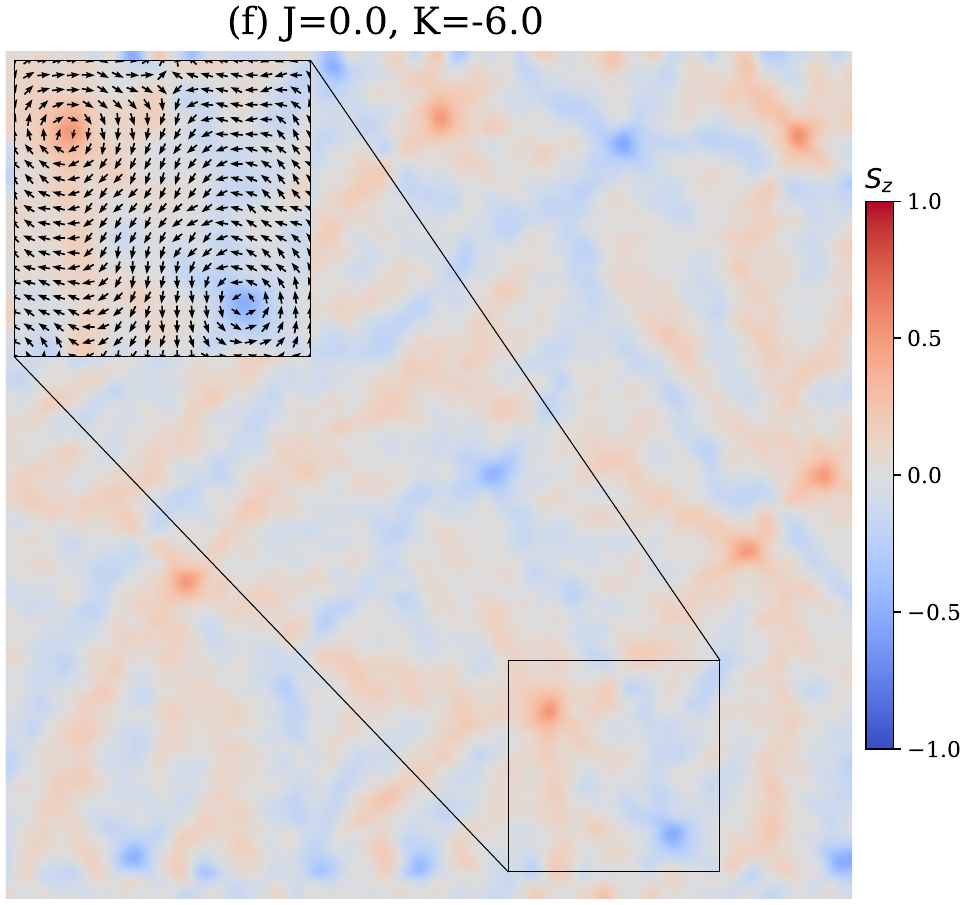}

\caption{Real-space spin configurations in the monolayer limit ($J=0.0$) for varying easy-plane anisotropy strengths ($K$). The color map represents the out-of-plane spin component $S_z$, while the arrows indicate the in-plane $(S_x, S_y)$ orientations. (a, b) At low anisotropies ($K=0.0$ and $K=-1.0$), the system stabilizes into topologically trivial single-$Q$ helical states. (c, d) Increasing the anisotropy to $K=-3.0$ and $K=-4.0$ triggers the nucleation of a highly ordered square MAX. (e) At $K=-5.0$, an extreme core-shrinking effect is observed, leading to a physical expansion of the inter-core distances and a ``softening'' of the topological crystal. (f) Under extreme anisotropy ($K=-6.0$), the macroscopic order collapses into a highly distorted, fragmented phase.}
\label{fig_1}
\end{figure}

\begin{figure}[!htbp]
    \centering
\includegraphics[width=7.4cm,height=7cm]{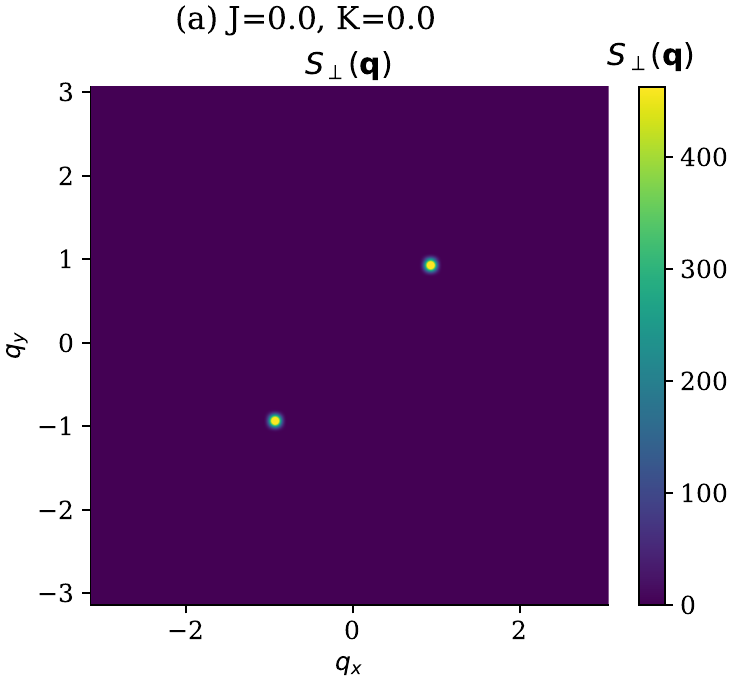}
\includegraphics[width=7.4cm,height=7cm]{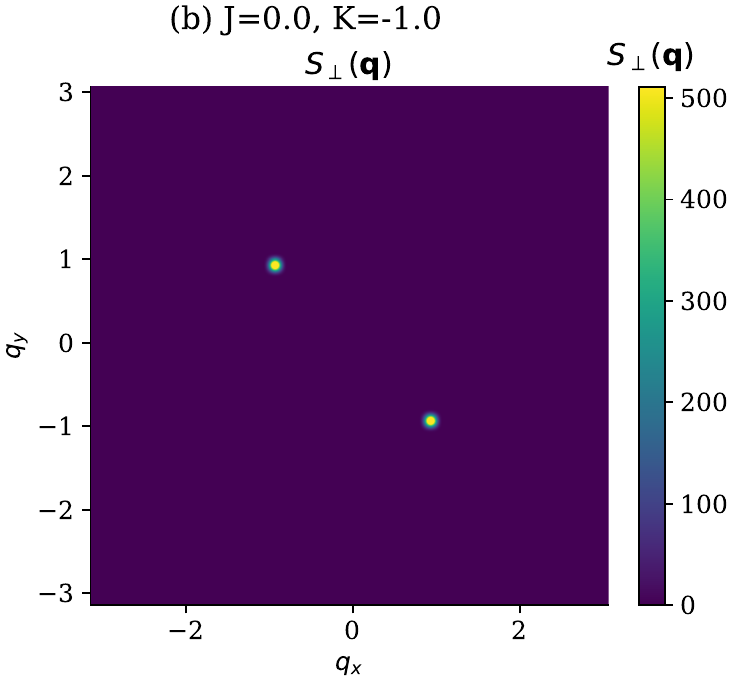}

\includegraphics[width=7.4cm,height=7cm]{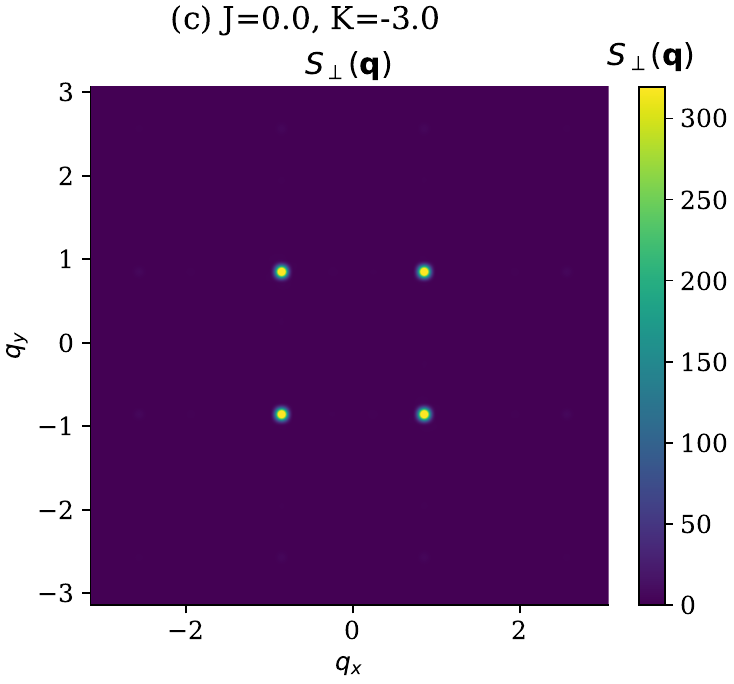}
\includegraphics[width=7.4cm,height=7cm]{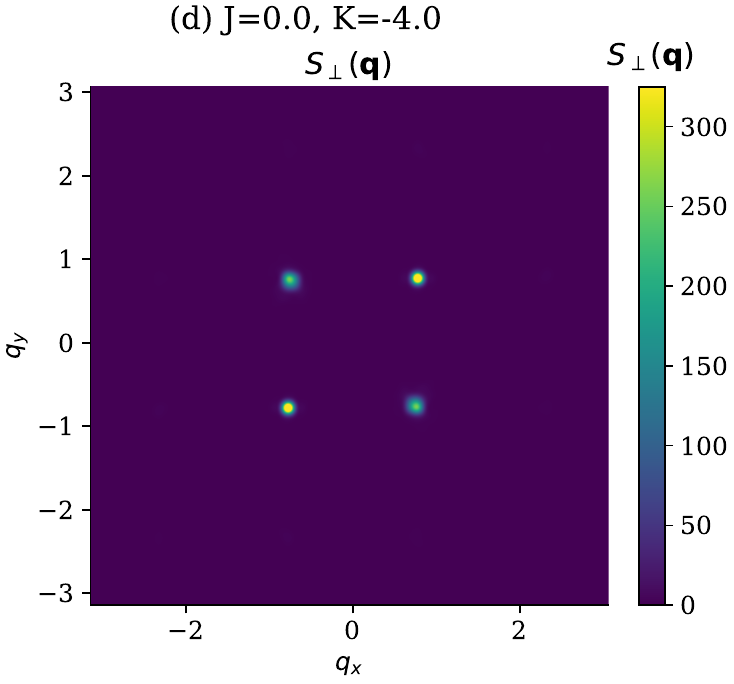}

\includegraphics[width=7.4cm,height=7cm]{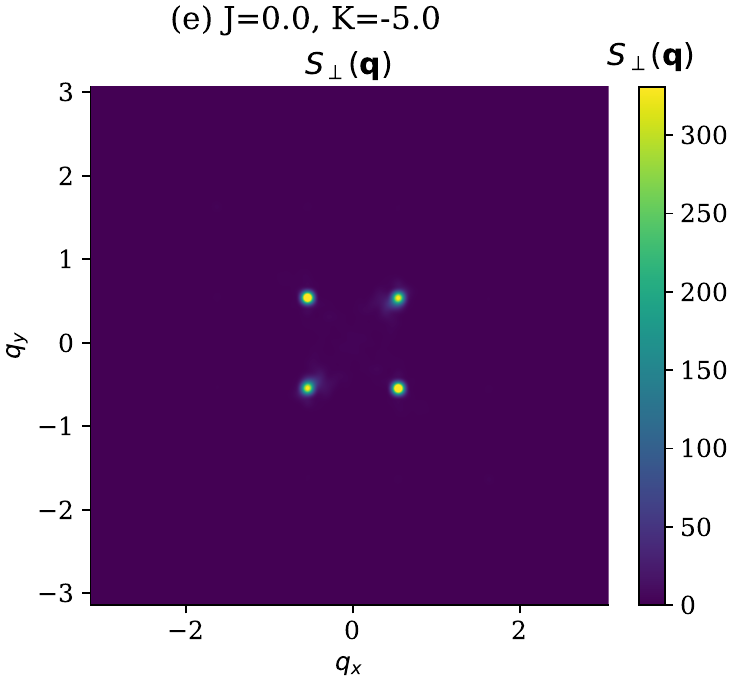}
\includegraphics[width=7.4cm,height=7cm]{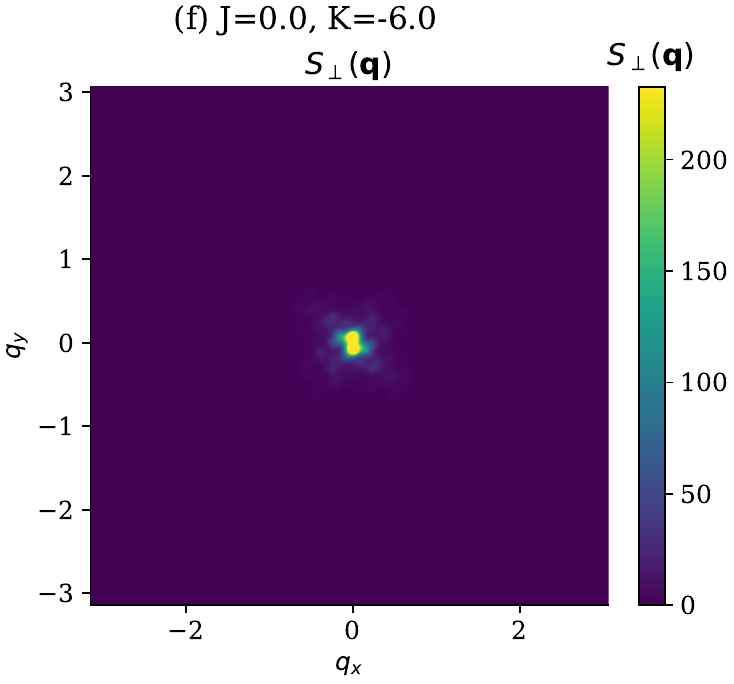}

\caption{Static spin structure factors $S_\perp(\mathbf{q})$ [Eq. (2)] corresponding to the uncoupled limit configurations ($J=0.0$) shown in Fig. \ref{fig_1}. (a, b) A single pair of Bragg peaks confirms the helical phase. (c) The four-fold symmetric peaks at $K=-3.0$ correspond to a MAX preserving $C_4$ rotational symmetry. (d, e) For $K=-4.0$ and $K=-5.0$, a pronounced intensity asymmetry emerges between the orthogonal peak pairs, revealing an anisotropic structural distortion and a precursor $C_4 \rightarrow C_2$ symmetry breaking before the onset of instability. (f) The transition to a diffuse central scattering pattern at $K=-6.0$ confirms the complete loss of long-range crystalline order.}
        \label{fig_2}
\end{figure}

In the uncoupled limit ($J=0$), the magnetic ground state is governed by the competition between DMI and the easy-plane anisotropy $K$. As shown in the real-space configurations of Fig. \ref{fig_1} and the corresponding structure factors in Fig.  \ref{fig_2}, the system undergoes a  topological phase transition when the $K$ changes.  For low anisotropy ($K=0.0, -1.0$), the system stabilizes in trivial helical (spiral) states. As $|K|$ increases to $-3.0$ and $-4.0$, we observe the nucleation of a square MAX. If the $|K|$ is further increased this macroscopic crystalline order is irrecoverably destroyed, and individual, isolated merons and antimerons emerge within the system \cite{ref_gao, ref_gobel}.

To verify the topological nature of these textures, we performed a local topological charge analysis using the Berg-Lüscher method (Eq. (3)). Although the detailed density maps are omitted here for brevity, our calculations consistently yield fractional charges of $|Q| \approx \pm 0.50 \pm 0.01$ per core, confirming that the emergent phase is indeed a MAX rather than an integer-charge skyrmion crystal.

A closer inspection of Figs. \ref{fig_1}(c-e) reveals that increasing $|K|$ leads to a significant shrinking of the meron cores. This core-shrinking effect reduces the effective coverage of the out-of-plane spin component ($S_z$), which in turn increases the relative inter-core distance within the MAX. This structural evolution is reflected in the reciprocal space (Fig. \ref{fig_2}); for $K=-3$, the structure factor displays four peaks of nearly equal intensity, consistent with ideal $C_4$ rotational symmetry. However, for $K=-4$ and $K=-5$, a distinct intensity asymmetry emerges, where one pair of Bragg peaks appears "dimmed" compared to the other.

This asymmetry indicates a spontaneous breaking of $C_4$ symmetry in favor of a distorted $C_2$ configuration. While traditional models of MAXs often assume perfect square symmetry, such anisotropy-induced distortions are a precursor to crystal instability.

To quantitatively assess the structural evolution of the MAX in the monolayer limit, we evaluate the real-space inter-core distance $d \approx 2\pi / |q^*|$ extracted from the fundamental Bragg peak positions $\mathbf{q}$ and corroborate these calculations with the real-space configurations shown in Fig. \ref{fig_1}. Here $|q^*|=\sqrt{q_x^{*2}+q_y^{*2}}$, where $q_x^*$ and $q_y^*$ are the peak position coordinates in reciprocal space shown in Fig. \ref{fig_2}.  As the easy-plane anisotropy increases from $K=-3.0$ to $K=-5.0$, the reciprocal peak coordinates ($|q_x^*|=|q_y^*|$) exhibit a substantial decrease from $0.864$ to $0.550$. This shift corresponds to a macroscopic physical expansion of the MAX constant from $d \approx 7.3$ to $d \approx 11.4$ lattice sites. Crucially, this reciprocal-space derivation is in perfect agreement with the direct observation of the defect density within the $L=80$ simulation domain (Fig. \ref{fig_1}); the number of topological cores per linear dimension drops from approximately 11 merons ($80/11 \approx 7.27$) for $K=-3.0$, to 10 merons ($80/10 = 8.0$) for $K=-4.0$, and further reduces to exactly 7 merons ($80/7 \approx 11.4$) for $K=-5.0$. This crystal expansion is a direct physical consequence of the extreme core-shrinking effect; as the out-of-plane domains are heavily penalized by the anisotropy, the in-plane winding regions expand, effectively "softening" the crystal pushing the topological cores further apart before the onset of structural instability.

\subsection{Spontaneous Topological Locking and Symmetry Restoration}
\label{subsec:locking}

\begin{figure}[!htbp]
    \centering
\includegraphics[width=4.3cm,height=4cm]{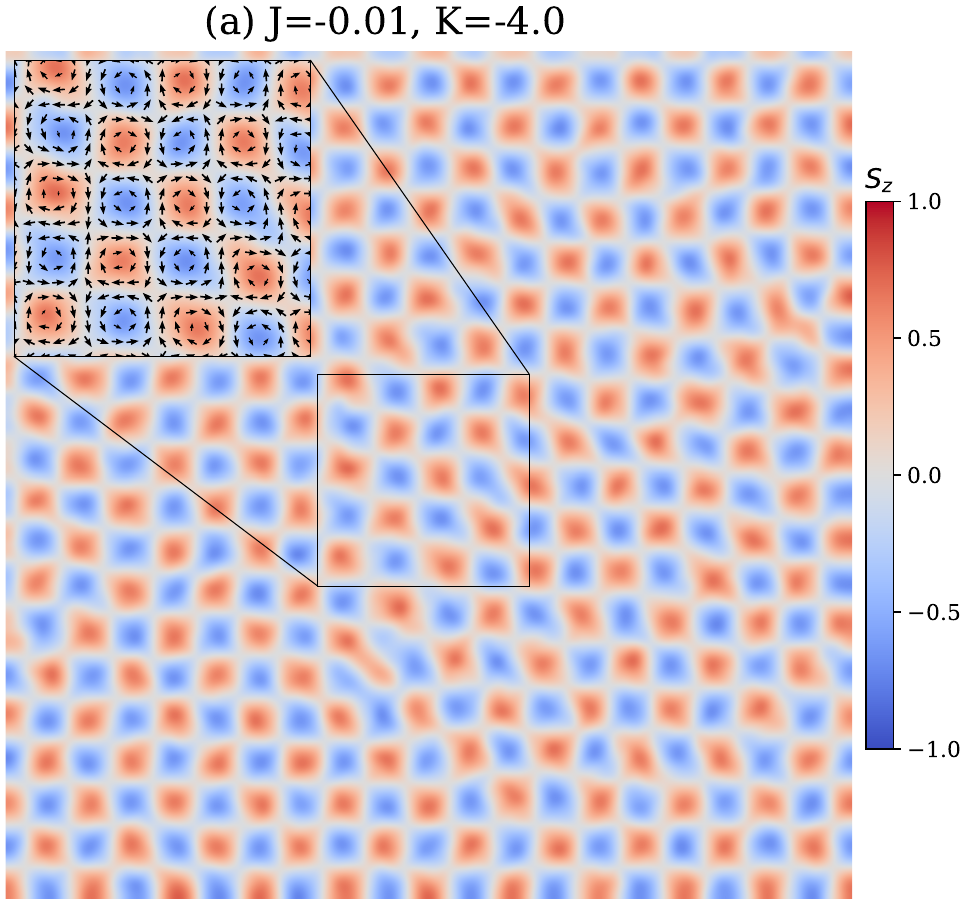}
\includegraphics[width=4.3cm,height=4cm]{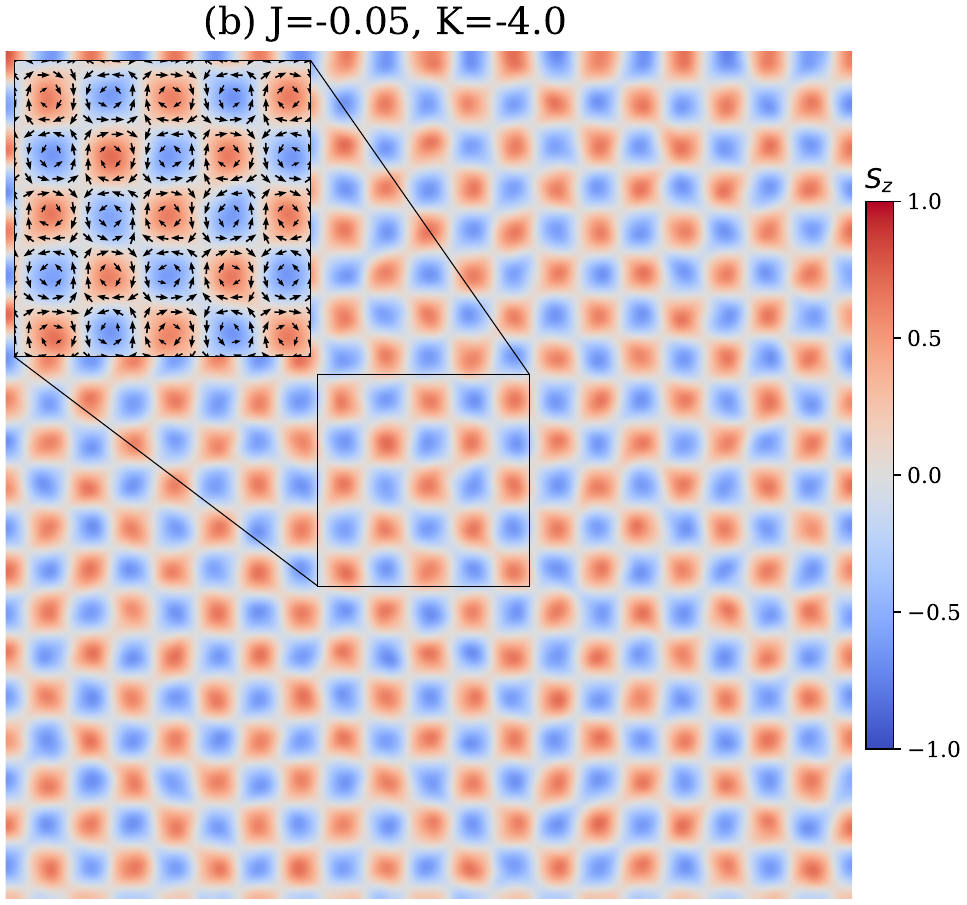}
\includegraphics[width=4.3cm,height=4cm]{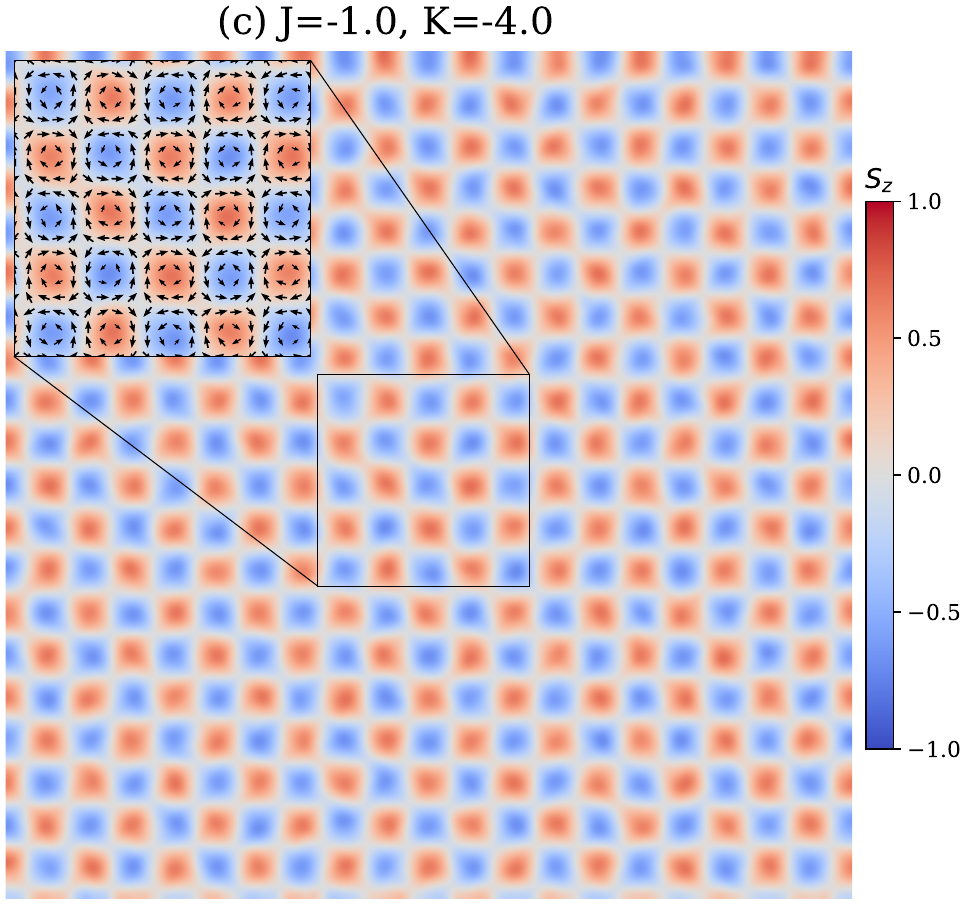}

\includegraphics[width=4.3cm,height=4cm]{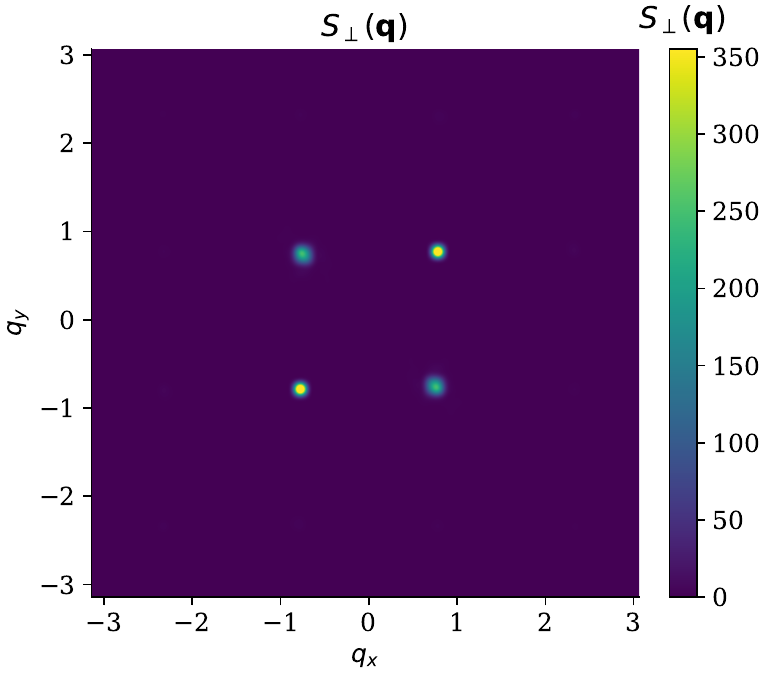}
\includegraphics[width=4.3cm,height=4cm]{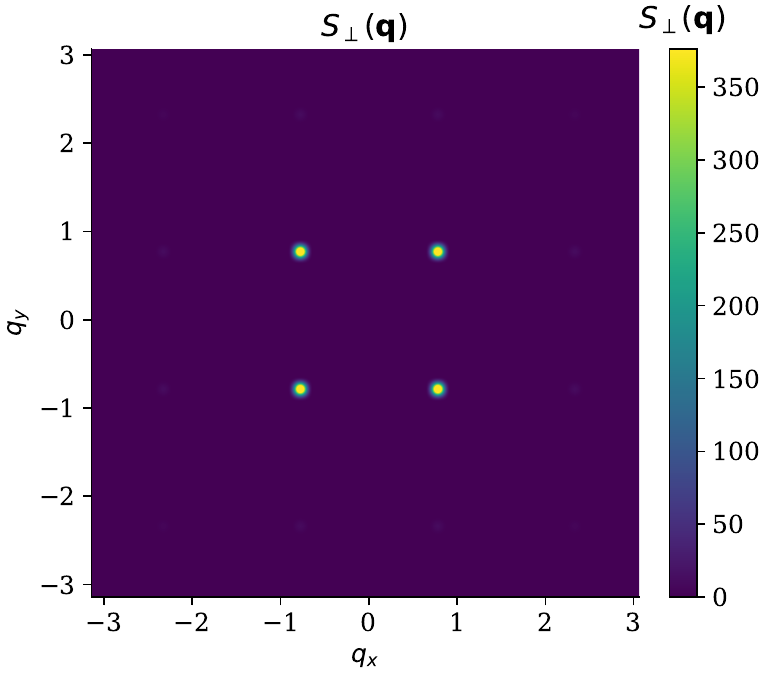}
\includegraphics[width=4.3cm,height=4cm]{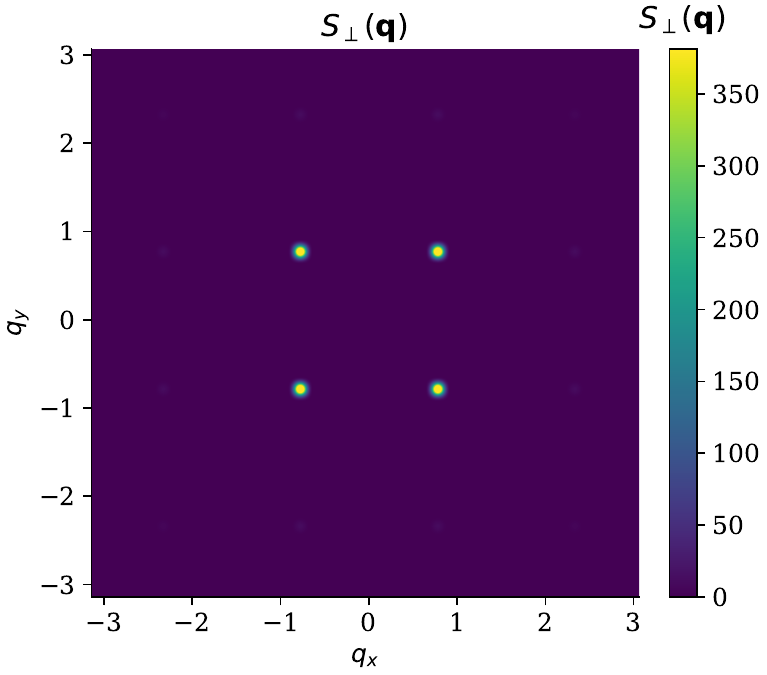}

\caption{Spontaneous structural symmetry restoration driven by interlayer exchange coupling for the $K=-4.0$ regime. The real-space configurations (top row) and corresponding structure factors (bottom row) are shown for (a) $J=-0.01$, (b) $J=-0.05$, and (c) $J=-1.0$. The introduction of the SAF coupling strictly equates the previously diverging orthogonal lattice spacings and perfectly equalizes the Bragg peak intensities. This indicates that the weak interlayer interaction acts as a structural scaffold, successfully restoring the global $C_4$ symmetry that was distorted in the monolayer limit.}
\label{fig_3}
\end{figure}

The introduction of interlayer exchange $J$ triggers the "topological locking" and "symmetry restoration" mechanisms. We first examine the representative case of $K=-4.0$ (Fig.  \ref{fig_3}). Although the real-space snapshots show subtle changes, the structure factor reveals a dramatic restoration: the previously dimmed peaks brighten and equalize, indicating that the $C_4$ symmetry is successfully restored. This structural regularization is further supported by the topological charge density maps in Fig.  \ref{fig_3}, showing a perfectly ordered MAX compared to the uncoupled state.

\begin{figure}[!htbp]
    \centering
\includegraphics[width=9.4cm,height=8cm]{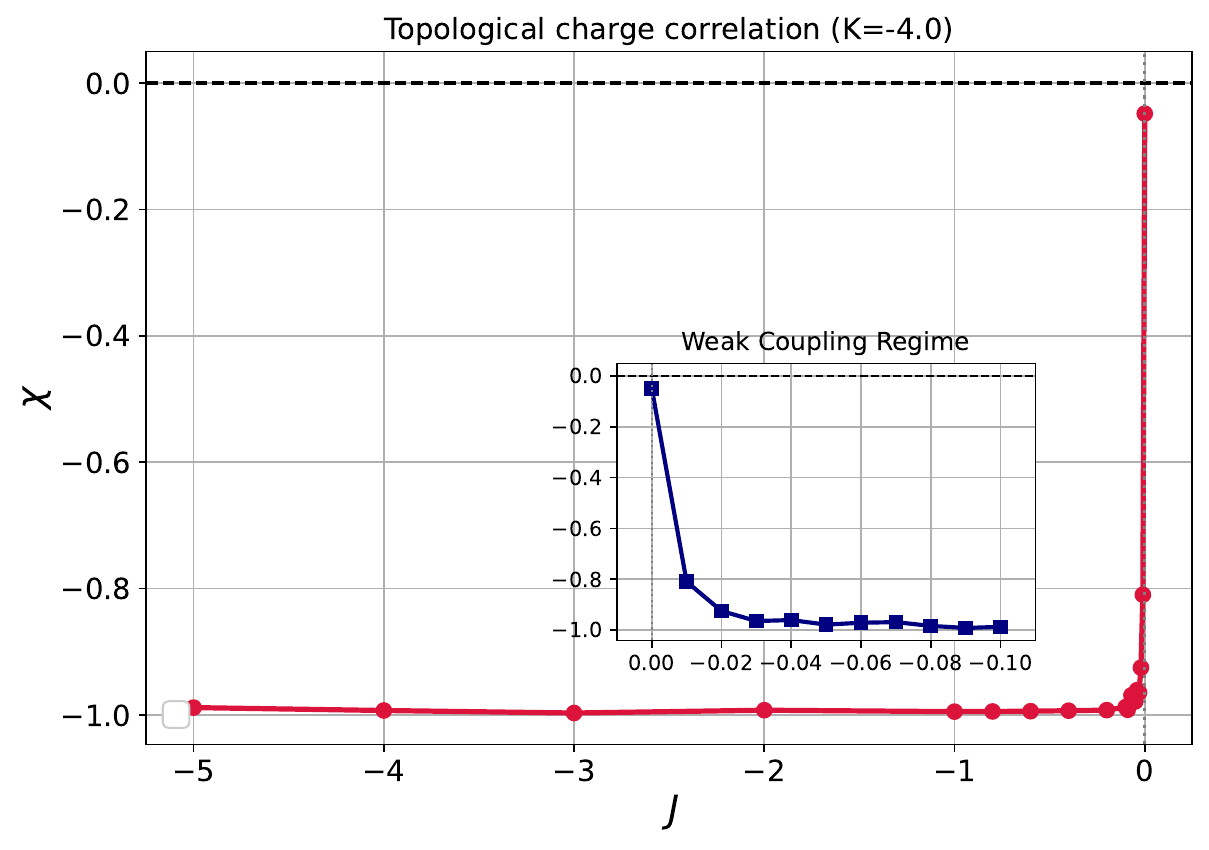}
\caption{Evolution of the interlayer topological charge correlation $\chi$ [Eq. (4)] as a function of the interlayer exchange coupling $J$ for $K=-4.0$. The main plot demonstrates the long-range saturation of the fully locked state. The inset provides a high-resolution view of the weak coupling regime ($J \in [-0.1, 0]$), revealing a sharp, first-order-like discontinuous jump at a critical threshold of $J_c \approx -0.01$. This abrupt transition physically marks the spontaneous vertical synchronization of the topological cores between the two layers.}
        \label{fig_4}
\end{figure}

\begin{figure}[!htbp]
    \centering
\includegraphics[width=9.4cm,height=8cm]{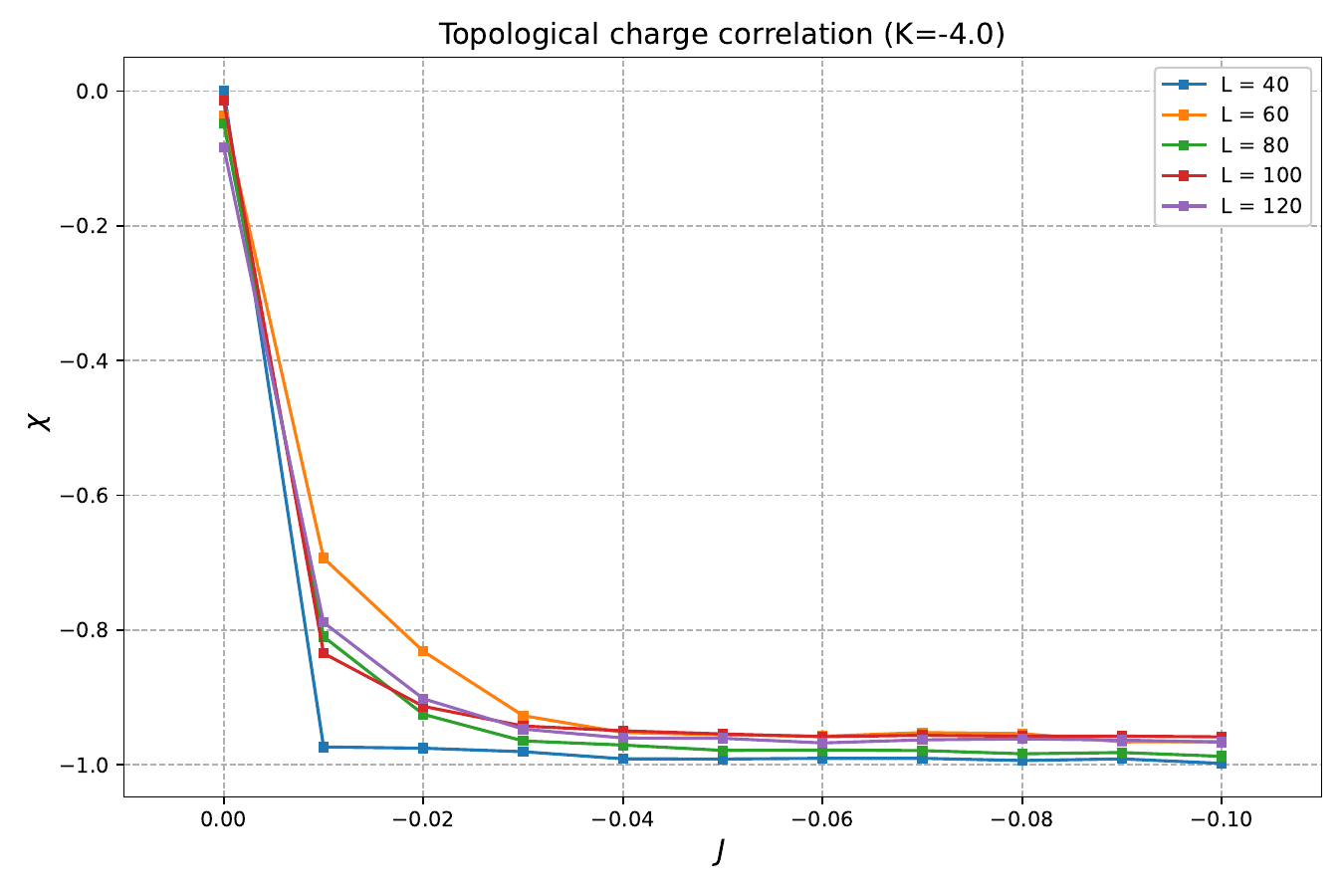}
\caption{Finite-size analysis of the topological locking transition for $K=-4.0$. The abrupt jump in the topological correlation at $J_c \approx -0.01$ remains highly consistent across varied simulation domain sizes from $L=40$ up to $120$. This size-independence verifies that the spontaneous topological synchronization is a robust, thermodynamically stable phase transition rather than a finite-size artifact.}
        \label{fig_5}
\end{figure}

This restoration is a direct consequence of "topological locking." As illustrated in Fig.   \ref{fig_4}, the interlayer correlation $\chi$ exhibits a sharp, discontinuous jump at an exceptionally weak threshold of $J_c \approx -0.01$, confirming the spontaneous nature of the locking. Our scaling analysis in Fig.  \ref{fig_5} shows that this jump is independent of the lattice size ($L=40$ to $120$), verifying its thermodynamic stability.

\begin{figure}[!htbp]
    \centering
\includegraphics[width=12.4cm]{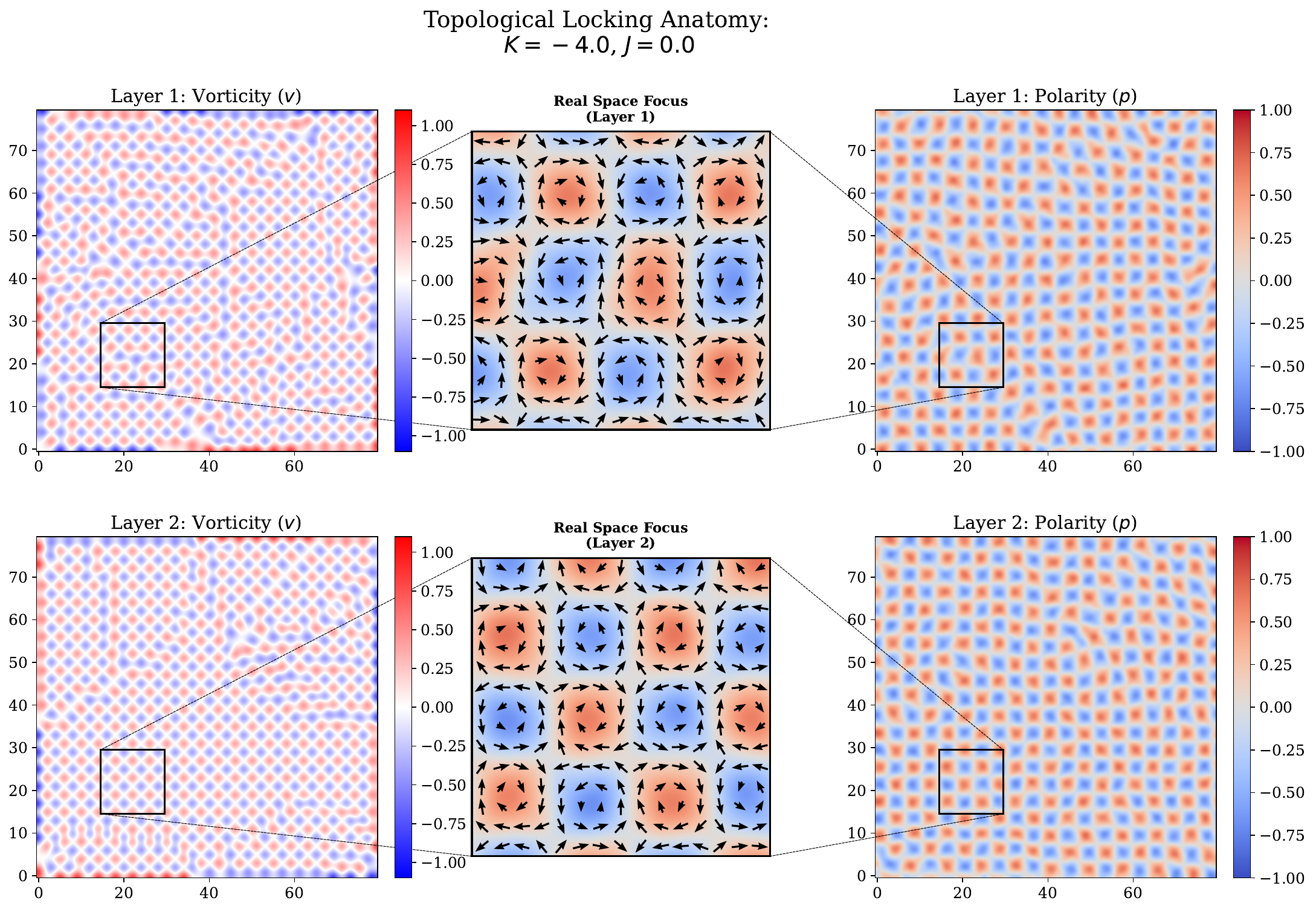}
\includegraphics[width=12.4cm]{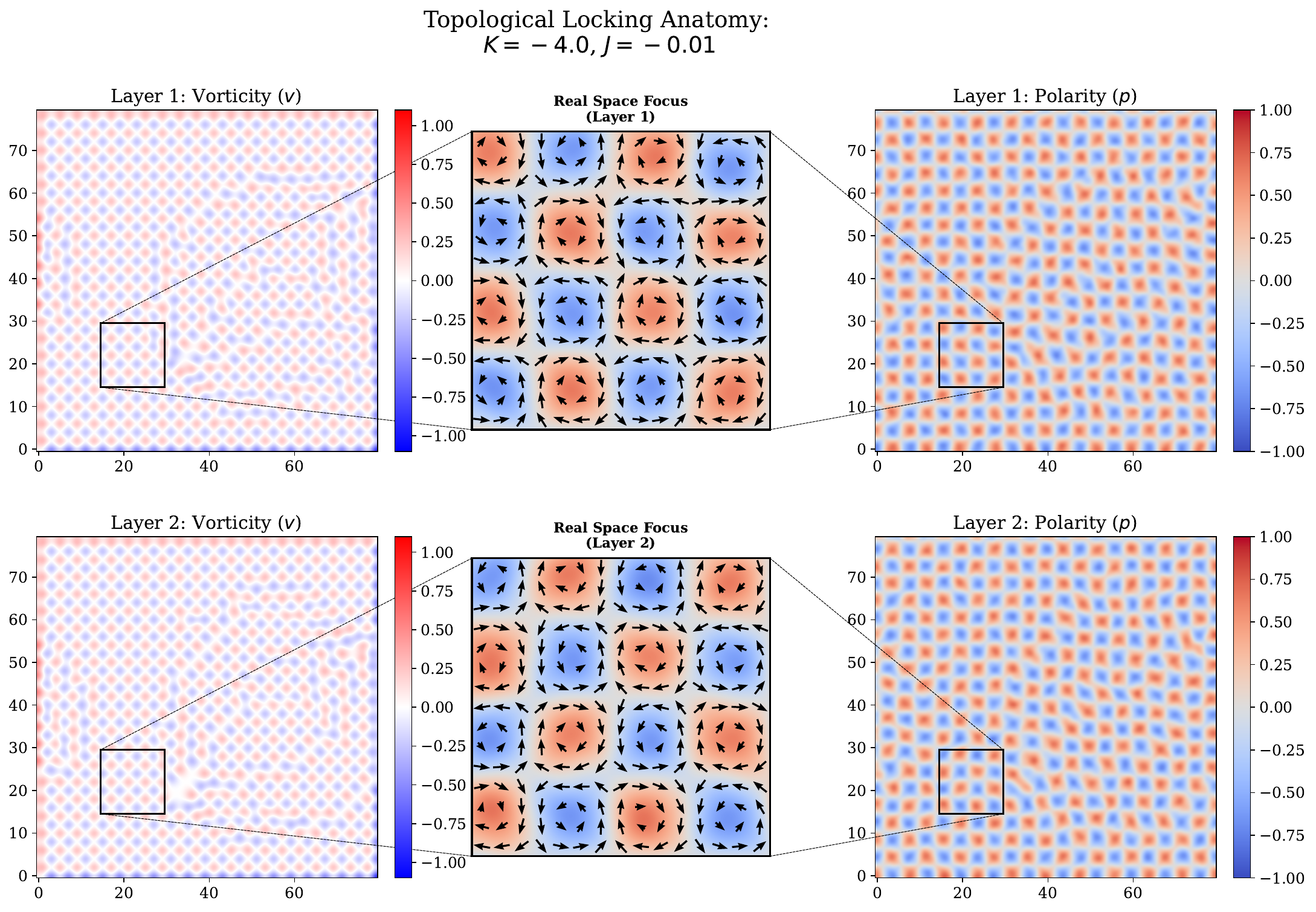}
\caption{Anatomy of the topological locking mechanism for $K=-4.0$. The left and right columns display the local vorticity ($v$) and polarity ($p \sim S_z$) maps, respectively. The central panels depict the magnified real-space focus of the corresponding layers. (a) In the uncoupled limit ($J=0.0$), the topological charges of Layer 1 and Layer 2 are vertically uncorrelated. (b) In the locked regime ($J=-0.01$), a strict one-to-one vertical correspondence is established. The fractional cores align vertically with strictly opposite polarities but identical vorticities, satisfying the relation $Q = p \cdot v / 2$ and resulting in the formation of robust antiferromagnetic bimeron dipoles.}
        \label{fig_6}

\end{figure}

To further characterize this state, we examine the core properties in Fig.  \ref{fig_6}. At $J=0$, the core polarities ($p$) and vorticities ($v$) of the two layers are uncorrelated, often resulting in a mismatch where a meron in the top layer may reside above an identical meron or empty space in the bottom layer. However, in the locked regime ($J < J_c$), a one-to-one correspondence is established: the meron (antimeron) cores align vertically such that their polarities are anti-parallel while their vorticities remain identical. Since the topological charge is defined as $Q = p v / 2$, this configuration results in the formation of robust AF-bimeron "topological dipoles" with opposite charges. This vertical synchronization acts as a structural scaffold, effectively "ironing out" the distortions observed in the monolayer limit. This mechanism remains qualitatively similar for $K=-3$ and $K=-5$.

\begin{figure}[!htbp]
    \centering
\includegraphics[width=12.4cm]{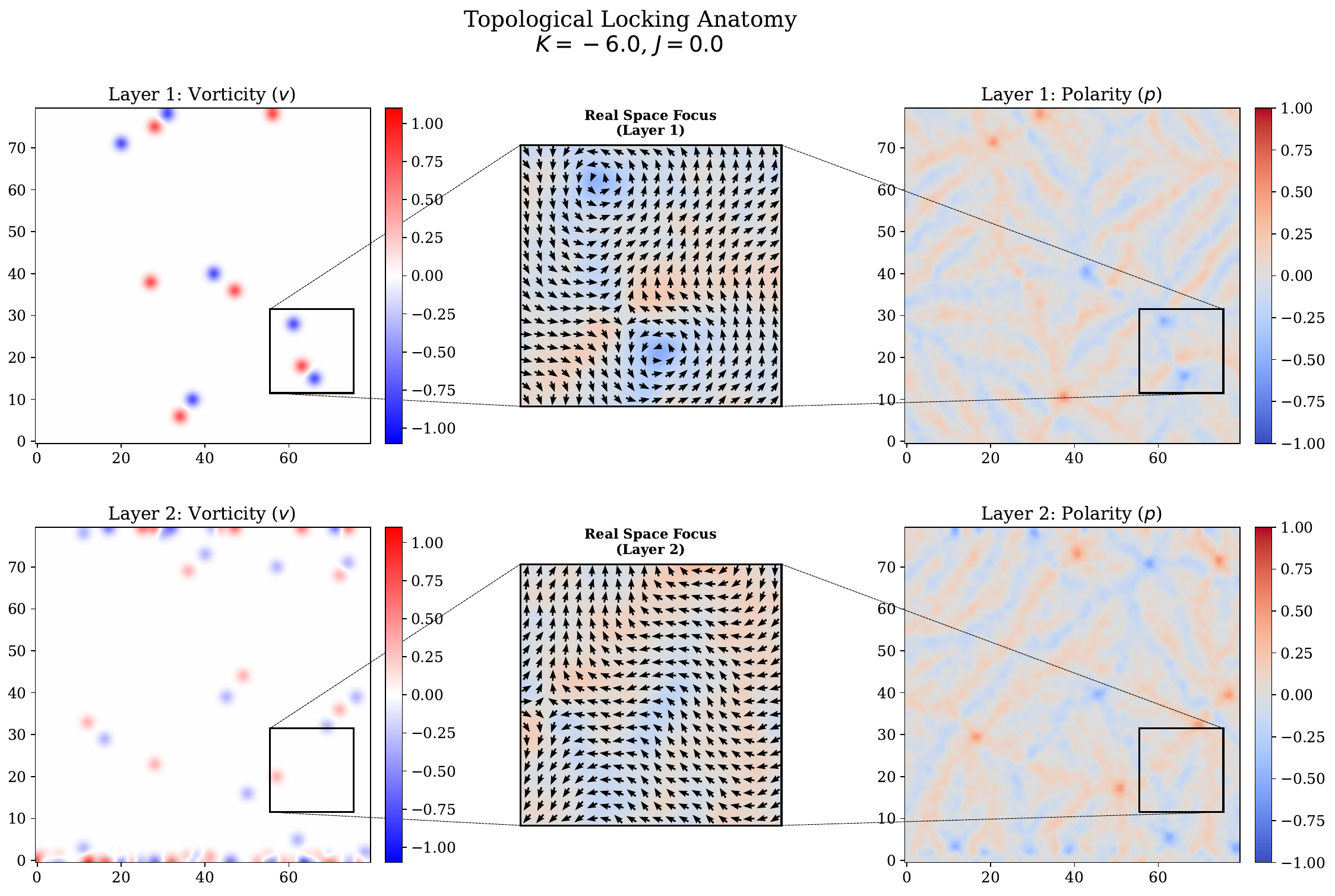}
\includegraphics[width=12.4cm]{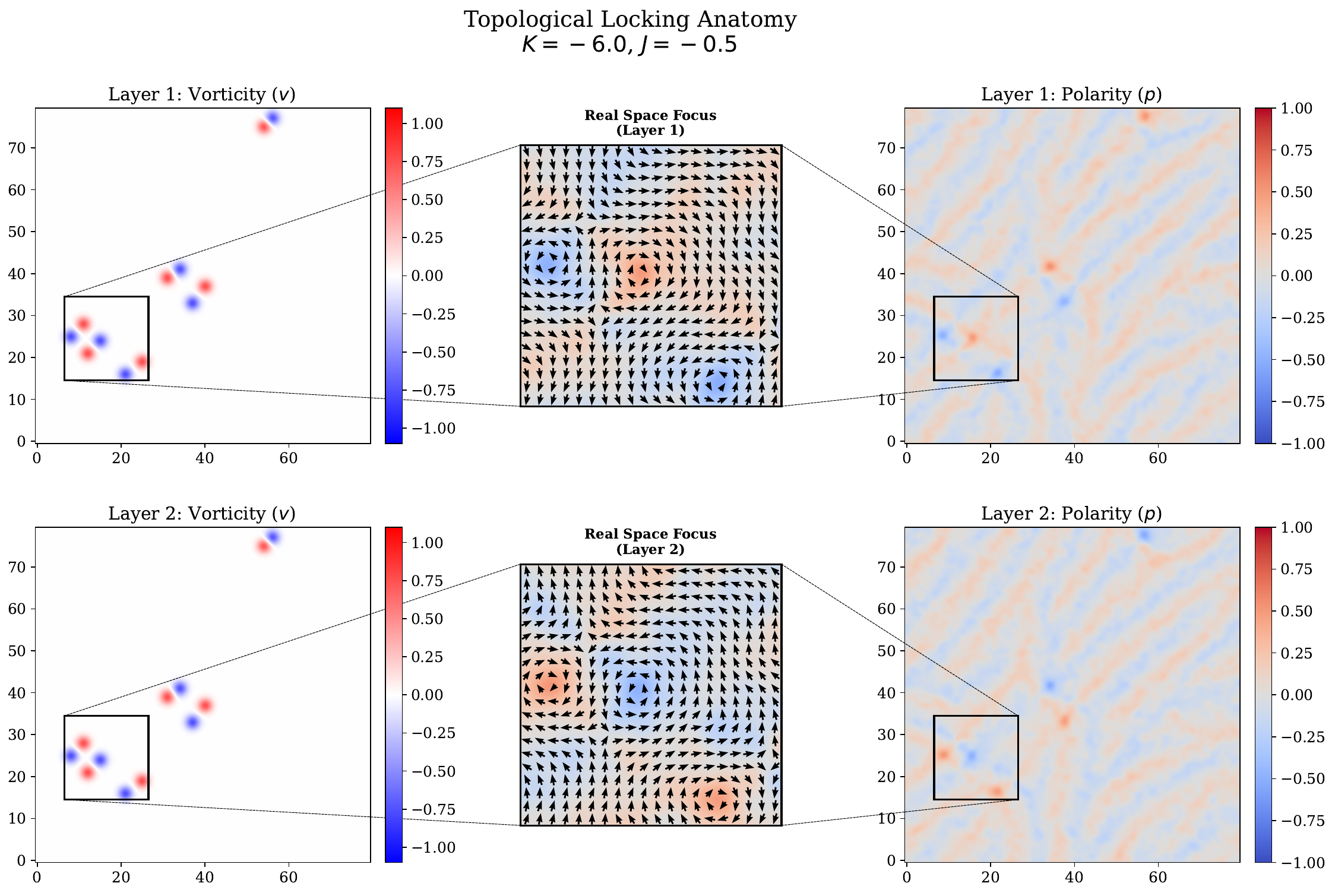}
\caption{The physical limits of the structural rescue mechanism under extreme easy-plane anisotropy at $K=-6.0$. (a) In the uncoupled state ($J=0.0$), the macroscopic crystal order is destroyed, leaving highly distorted and irregularly distributed defects. (b) Under a significantly stronger interlayer coupling ($J=-0.5$), the system fails to resurrect the global $C_4$ symmetric crystal structure. However, the strong SAF coupling successfully enforces a strict \textit{local} topological locking, actively aligning the surviving, fragmented defects into isolated AF-bimeron dipoles. This configuration highlights the critical decoupling between local vertical synchronization and global structural order in the melted phase limit.}
        \label{fig_7}

\end{figure}

Finally, we examine the limits of the structural rescue mechanism under extreme easy-plane anisotropy at $K=-6.0$ (Fig. \ref{fig_7}). In this regime, the uncoupled system ($J=0$) loses its long-range crystalline order, and the MAX completely decays into a highly distorted, fragmented phase. Consequently, the exceptionally weak coupling ($J_c \approx -0.01$) that successfully regularized the $K=-4.0$ system is entirely insufficient to induce any inter-layer synchronization.

Interestingly, when a substantially stronger interlayer antiferromagnetic exchange is applied ($J = -0.5$), the system still fails to restore the global $C_4$ symmetry, confirming that the symmetry restoration of the macroscopic lattice has a definitive upper physical bound. However, as demonstrated in the topological locking anatomy of Fig. \ref{fig_7}, this strong SAF coupling successfully enforces a strict local topological locking. The surviving, irregularly distributed defects vertically align to form isolated AF-bimeron dipoles, preserving the $Q = p v / 2$ relationship with opposite polarities and identical vorticities. This reveals a critical decoupling between local synchronization and global structural order: while strong interlayer exchange can locally lock topological charges even within a "melted" phase, it cannot resurrect a macroscopic lattice once the anisotropy-induced structural decay surpasses a critical threshold.

The impact of the interlayer exchange $J$ on this crystal spacing reveals two distinct dynamical regimes depending on the underlying crystal stiffness. For moderate anisotropies ($K=-3.0$ and $-4.0$), the peak positions remain remarkably invariant ($|q_x^*|=|q_y^*| \approx 0.864$ and $0.785$, respectively) across all values of $J$. In these rigid crystal regimes, the weak coupling strictly enforces the vertical topological locking and $C_4$ symmetry restoration without physically altering the native lattice constant dictated by the DMI.

However, a profoundly different behavior emerges for the highly expanded and "softened" crystal at $K=-5.0$. When the interlayer coupling exceeds a critical threshold ($J \le -0.05$), the Bragg peaks abruptly shift from $|q_x^*|=|q_y^*|= = 0.550$ to $|q_x^*|=|q_y^*| = 0.628$. This transition corresponds to an anomalous interlayer-induced lattice compression, where the macroscopic crystal contracts from $d \approx 11.4$ to $d \approx 10.0$. In this pre-collapse regime, the MAX is sufficiently flexible that the strong SAF coupling pulls the fractional cores closer together, actively compressing the crystal to maximize the vertical antiferromagnetic exchange energy just prior to the complete structural breakdown observed at $K=-6.0$.

\section{Conclusion}\label{sec:conclusion}

In summary, we have systematically investigated the spontaneous topological locking and structural symmetry restoration of MAXs in SAF bilayers using large-scale Monte Carlo simulations. Our findings demonstrate that SAF architectures serve not merely as passive hosts for topological textures, but as active structural scaffolds capable of rescuing fractional spin configurations from anisotropy-induced degradation.

In the uncoupled monolayer limit, the interplay between the DMI and strong easy-plane anisotropy strictly dictates the structural integrity of the MAX. We quantitatively showed that increasing the anisotropy induces an extreme core-shrinking effect, which physically expands the inter-core distance and triggers a precursor $C_4 \rightarrow C_2$ symmetry breaking prior to a complete structural collapse. However, the introduction of an exceptionally weak interlayer antiferromagnetic exchange ($J_c \approx -0.01$) fundamentally alters this thermodynamic trajectory. This ultra-weak coupling is sufficient to trigger a sharp, spontaneous vertical synchronization, perfectly aligning the fractional cores into robust antiferromagnetic bimeron dipoles satisfying $Q = p v / 2$. For structurally rigid crystals, this local locking mechanism seamlessly translates into global structural repair, ironing out the spatial distortions and fully restoring the macroscopic $C_4$ rotational symmetry.

Furthermore, our reciprocal-space analysis reveals the nuanced dependency of this topological rescue on the underlying stiffness of the crystal. For highly expanded and softened crystals nearing the instability threshold ($K=-5.0$), the SAF coupling actively induces an anomalous macroscopic compression to maximize the interlayer exchange energy. Probing the extreme anisotropy limit ($K=-6.0$) exposed a definitive physical boundary: while the macroscopic crystalline order cannot be resurrected once fully decayed, strong interlayer coupling still enforces a strict local topological locking of the surviving isolated defects. This fundamental decoupling between local synchronization and global structural order highlights both the robust resilience and the definitive physical limits of the SAF stabilization mechanism.

Ultimately, these results provide a comprehensive theoretical framework for the stabilization of fractional topological textures in complex magnetic environments. By utilizing interlayer exchange coupling to actively suppress structural distortions and tightly bind these fractional states, SAF architectures emerge as a highly robust, scalable platform for high-density, Magnus-force-free topological information processing.

\end{document}